\documentclass[11pt,a4paper]{article}
\pdfoutput=1

\usepackage{epsfig}
\usepackage{amssymb}
\usepackage{graphicx}
\usepackage{color}
\usepackage{subfigure}
\usepackage{mathtools}
\usepackage[hidelinks]{hyperref}

\makeatletter
\renewcommand{\theequation}{\thesection.\arabic{equation}}
\@addtoreset{equation}{section} \makeatother
\renewcommand{\vec}[1]{\underline{#1}}
\newcommand{\lra}[1]{\la{#1}\ra}

\newcommand{\sss}[1]{\scriptscriptstyle{#1}}

\def\a{\alpha}

\def\d{\delta}
\def\D{\Delta}

\def\e{\eta}
\def\ph{\phi}
\def\Ph{\Phi}

\def\l{\lambda}
\def\L{\Lambda}
\def\m{\mu}

\def\th{\theta}

\def\s{\sigma}
\def\S{\Sigma}
\def\ta{\tau}

\def\lt{\left}
\def\rt{\right}

\def\nn{\nonumber}

\DeclareMathOperator{\tr}{Tr}

\def\p{\partial}

\def\la{\langle}
\def\ra{\rangle}

\def\wt{\widetilde}

\def\nn{\nonumber}

\def\larw{\leftarrow}

\def\bea{\begin{eqnarray}}
\def\eea{\end{eqnarray}}

\def\nf{{\sss{N_F}}}

\begin{document}

\begin{titlepage}
\title{\vskip -60pt
\vskip 20pt %
Junctions of mass-deformed nonlinear sigma models on the Grassmann manifold
}
\author{
Sunyoung Shin \footnote{e-mail : sihnsy@skku.edu}}
\date{}
\maketitle \vspace{-1.0cm}
\begin{center}
~~~
\it  Institute of Basic Science, Sungkyunkwan University, \\
Suwon 16419, Republic of Korea

\end{center}

\thispagestyle{empty}

\begin{abstract}
We study vacua and walls of the mass-deformed nonlinear sigma models on the Grassmann manifold $G_{N_F,N_C}=\frac{SU(N_F)}{SU(N_C)\times SU(N_F-N_C)\times U(1)}$ and discuss three-pronged junctions for $N_C=1,2,3$ in four dimensions.
\end{abstract}

\end{titlepage}
\section{Introduction} 
\setcounter{equation}{0}

Topological defects play important roles in model building. Our world can be realized on extended topological defects such as walls or junctions in brane world scenarios \cite{ArkaniHamed:1998rs,Randall:1999ee}. Wall networks are applied to study dark matter and dark energy \cite{Bucher:1998mh}.

Walls interpolating isolated supersymmetric vacua preserve $1/2$ supersymmetry. Walls in Abelian gauge theories are studied in \cite{Abraham:1992vb,Tong:2002hi}. The moduli matrix formalism is proposed to analyse walls of ${\mathcal{N}}=2$ non-Abelian gauge theories \cite{Isozumi:2004jc}. In the strong coupling limit, the model with $N_F>N_C$ becomes the mass-deformed hyper-K\"{a}hler nonlinear sigma model on $T^\ast G_{N_F,N_C}$. In this limit, the wall solutions become exact. It is shown that with the Fayet-Iliopoulos parameters $(b,b^\ast,c)=(0,0,c>0)$, we can find a bundle structure so that vacua and walls are on the K\"{a}hler manifold, which can be realized as the ${\mathcal{N}}=1$ non-Abelian gauge theory. The moduli matrix formalism is thoroughly examined in the mass-deformed nonlinear sigma models on the Grassmann manifold, which produce exact Bogomol'nyi-Prasad-Sommerfield (BPS) solutions \cite{Isozumi:2004jc}.

We can consider models with superpotentials as vacua and walls of the mass-deformed nonlinear sigma models can be analytically analysed on the Grassmann manifold. The simplest models that have exact BPS solutions are the mass-deformed nonlinear sigma models on the quadrics of the Grassmann manifold $SO(2N)/U(N)$ and $Sp(N)/U(N)$. The nonlinear sigma models on $SO(2N)/U(N)$ and $Sp(N)/U(N)$  are constructed as gauge theories by (anti-)holomorphic embeddings to the Grassmann manifold \cite{Higashijima:1999ki}. By integrating out the Lagrange multipliers of the superpotential terms, we obtain the F-term constraints. By introducing mass-terms to the models, the most part of the continuous vacuum is lifted and a finite number of discrete vacua remain on the surfaces. We can find exact BPS solutions that interpolate the discrete vacua.

In the moduli matrix formalism, walls are algebraically constructed from elementary walls. The elementary walls can be identified with simple roots of the global symmetry \cite{Sakai:2005sp}. In \cite{Lee:2017kaj}, a pictorial representation is proposed to study vacua and walls of mass-deformed K\"{a}hler nonlinear sigma models on $SO(2N)/U(N)$ and $Sp(N)/U(N)$, which are quadrics of the Grassmann manifold $G_{2N,N}$. It is observed that we can produce the whole structure of vacua and elementary walls from the vacuum structures that are connected to the maximum number of elementary walls for generic $N$. There is a recurrence of a two dimensional diagram for each $N$ mod 4 in the vacuum structures on $SO(2N)/U(N)$ and $Sp(N)/U(N)$. The vacuum structures are proved by induction.

The observation made in \cite{Lee:2017kaj} should be useful not only for constructing BPS walls, but also for analysing supersymmetric models, which have 1/1 vacua, 1/2 BPS objects and 1/4 BPS objects since the moduli matrix formalism produces exact solutions in the mass-deformed nonlinear sigma models on  the compact Hermitian symmetric spaces of the quotient with respect to the gauge group.

There are two standard off-shell ${\mathcal{N}}=2$ superspace formalisms. One is the harmonic superspace formalism and the other is the projective superspace formalism. In the harmonic superspace formalism, the nonlinear sigma models on the complex projective spaces and on the Grassmann manifold are constructed as the quotient with respect to the gauge group \cite{Galperin:1985dw}, but to the best of the author's knowledge, it is not known whether (anti-)holomorphic embeddings can be defined. In the projective superspace formalism, the holomorphic embeddings for the nonlinear sigma models on the compact Hermitian symmetric spaces are well-defined but the models are constructed in the inhomogeneous coordinates of the target space \cite{Arai:2006gg}. 
Therefore provided that we are willing to construct models of the quotient with respect to the gauge group, it is useful to study vacua and BPS objects of the ${\mathcal{N}}=2$ mass-deformed nonlinear sigma models on the quadrics of the Grassmann manifold in the ${\mathcal{N}}=1$ superspace formalism to get better understanding of the ${\mathcal{N}}=2$ supersymmetric model building.

Intersecting walls form junctions, which preserve $1/4$ supersymmetry \cite{Abraham:1990nz,Gibbons:1999np,Carroll:1999wr,Saffin:1999au,Bazeia:1999xi}. The energy density of a wall junction is bounded from below by central charge densities ${\mathcal{Z}}_\a$ $(\a=1,2)$ and ${\mathcal{Y}}$. ${\mathcal{Z}}_\a$ are components of the tension vector of the wall, which pulls the junction along the wall direction outwards and ${\mathcal{Y}}$ is the charge density for the junction. An Abelian junction has a negative $Y$-charge while a non-Abelian junction has a positive $Y$-charge. In the strong coupling limit, the charge density ${\mathcal{Y}}$ vanishes.

A three-pronged junction is formed by a set of three vacua that are interpolated by walls. Three-pronged
junction solutions are discussed in the moduli matrix formalism \cite{Eto:2005cp,Eto:2005fm,Eto:2007uc}. The moduli matrix
formalism can be easily applied to Abelian junctions, but it becomes complicated with non-Abelian junctions. In
\cite{Eto:2005fm}, three-pronged junctions of mass-deformed nonlinear sigma models on the Grassmann manifold $G_{N_F,N_C}$ are analyzed
by the Pl\"{u}cker embedding. The models are embedded into the complex projective space $\mathbf{C}P^{_{N_F}C_{N_C}-1}$, which
produce Abelian gauge theories. The Pl\"{u}cker embedding is useful for the Grassmann manifold, but the drawback of it is that the method is not directly applicable to the quadrics of the Grassmann manifold.

An alternative method is proposed to construct three-pronged junctions of the mass-deformed nonlinear sigma models on the Grassmann manifold \cite{Shin:2018chr}. The pictorial representation, which is introduced by \cite{Lee:2017kaj}, is applied to vacua and walls of mass-deformed nonlinear sigma models on the Grassmann manifold. The vacua and the walls of the mass-deformed nonlinear sigma models on $G_{N_F,N_C}$
with $(N_F,N_C)=(4,2)$,$(5,2)$,$(5,3)$,$(6,2)$, $(6,3)$,$(6,4)$ are explicitly shown in the representation. In the representation, the duality of the Grassmann manifold $N_C \leftrightarrow N_F-N_C$ is realized as a $\pi$-rotation of the diagram. The diagram of $G_{N_F+1,N_C}$ repeats the diagram of $G_{N_F,N_C}$. It is also shown in \cite{Shin:2018chr} that by reformulating the diagrams in the pictorial representations, we can produce polyhedra, which are similar to the polyhedra that are introduced to study BPS objects of the mass-deformed nonlinear sigma models on the complex projective space \cite{Eto:2005cp}. Vertices, edges and triangular faces correspond to vacua, walls and three-pronged junctions. Positions of three-pronged junctions are computed by making use of the polyhedra instead of using the Pl\"{u}ker embedding \cite{Shin:2018chr}.

In this paper, we discuss vacua, walls and three-pronged junctions of mass-deformed nonlinear sigma models on the Grassmann manifold with flavour number $N_F$ and colour number $N_C$$=$$1$,$2$,$3$. Since diagrams in the pictorial representation consist of the simple roots of $SU(N_F)$, diagrams for $N_C\geq 4$ are trivial extensions.

This paper is organized as follows. In Section \ref{sec:model}, we review the model \cite{Lindstrom:1983rt,Arai:2003tc} and the moduli matrix formalism \cite{Isozumi:2004jc}. In Section \ref{sec:vac_walls}, we review the work \cite{Shin:2018chr} and discuss vacua and walls of the mass-deformed nonlinear sigma models on $G_{N_F,N_C}$ with $N_C=1,2,3$. In Section \ref{sec:junctions}, we apply the method, which is introduced by \cite{Shin:2018chr}, to three-pronged junctions of the mass-deformed nonlinear sigma model on $G_{N_F,3}$. In Section \ref{sec:summary}, we summarize our results. In \ref{app:g84}, we present the diagram for $G_{8,4}$ to show that the method discussed in Section \ref{sec:vac_walls} is generically applicable.

\section{Model}  \label{sec:model}
\setcounter{equation}{0}
We study junctions of the $\mathcal{N}=2$ mass-deformed nonlinear sigma model on the Grassmann manifold $G_{N_F,N_C}=\frac{SU(N_F)}{SU(N_C)\times SU(N_F-N_C) \times U(1)}$ in four dimensions. The Lagrangian of the mass-deformed nonlinear sigma model \cite{Lindstrom:1983rt,Arai:2003tc} is
\begin{align} \label{eq:n=2_lag}
{\mathcal{L}}=&\int d^4\th \tr\lt[\Ph\Ph^\dagger e^V +\Psi^\dagger \Psi e^{-V} - c V\rt] \nn\\
&+\int d^2\th \tr \lt[ \L \lt(\Ph\Psi-b I_M\rt)+\Ph {\mathcal{M}} \Psi+\mathrm{(conjugate~transpose)} \rt],\nn\\
&( c \in \mathbf{R}_{\geq 0},~b\in \mathbf{C}),
\end{align}
where $c$ is the electric Fayet-Iliopoulos (FI) parameter and $b$, $b^\ast$ are the magnetic FI parameters. Chiral fields $\Ph_a^{~i}(x,\th,\bar{\th})$, $\Psi_i^{~a}(x,\th,\bar{\th})$, $\L_a^{~b}$ and vector field $V_a^{~b}(x,\th,\bar{\th})$ are matrix valued and defined as follows:
\begin{align} \label{eq:comp_fld}
&\Ph_a^{~i}(y)=\ph_a^{~i}(y)+\sqrt{2}\th\psi_a^{~i}(y)+\th\th F_a^{~i}(y),
~~\lt(y^\m=x^\m+i\th\s^\m\bar{\th}\rt), \nn\\
&\Psi_i^{~a}(y)=\varphi_i^{~a}(y)+\sqrt{2}\th\chi_i^{~a}(y)+\th\th G_i^{~a}(y), \nn\\
&V_a^{~b}(x)=2\th\s^\m\bar{\th}A_{\m a}^{~~b}(x)+i\th\th\bar{\th}\bar{\l}_a^{~b}(x)-i\bar{\th}\bar{\th}\th\l_a^{~b}(x)
+\th\th\bar{\th}\bar{\th}D_a^{~b}(x), \nn\\
&\L_a^{~b}(y)=-{\mathcal{S}}_a^{~b}(y)+\th\e_a^{~b}(y)+\th\th K_a^{~b}(y), \nn\\
&(a=1,\cdots,N_C;~i=1,\cdots,N_F;~\m=0,\cdots,3).
\end{align}
We diagonalise $V$ and $\L$ for later use.

The $SU(N)$ Cartan generators $\vec{H}=(H_1,\cdots,H_N)$ \cite{Isaev:2018xcg}
are defined by
\bea \label{eq:cartan_su_n}
H_n=e_{n,n}-\frac{1}{N}I_{N\times N}, \quad (n=1,\cdots,N),
\eea
where $e_{p,q}$ is an $N\times N$ matrix of which the $(p,q)$ component is one.
The complex mass matrix ${\mathcal{M}}$ can be formulated as a linear combination of (\ref{eq:cartan_su_n}) with complex parameters. The mass matrix ${\mathcal{M}}$ is a traceless diagonal matrix.

The Lagrangian (\ref{eq:n=2_lag}) with the component fields (\ref{eq:comp_fld}) can be computed. The equations of the auxiliary fields are solved by
\begin{align}
&F=-\varphi^\dagger{\mathcal{M}}^\dagger+{\mathcal{S}}^\dagger\varphi^\dagger,\quad\mbox{(conjugate~transpose)},\nn\\
&G=-{\mathcal{M}}^\dagger\ph^\dagger+\ph^\dagger{\mathcal{S}}^\dagger,\quad\mbox{(conjugate~transpose)}.
\end{align}
Then the bosonic part of the Lagrangian is
\begin{align}\label{eq:lag0}
{\mathcal{L}}=\tr\Big[&D_\mu\ph({D^\m\ph})^\dagger + (D_\mu\varphi)^\dagger D^\mu\varphi 
-|\ph {\mathcal{M}}-{\mathcal{S}}\ph|^2-| {\mathcal{M}}\varphi-\varphi {\mathcal{S}}|^2 \nn\\
&+(\ph\ph^\dagger-\varphi^\dagger\varphi-cI_{\sss{N_C}})D +K(\ph\varphi-bI_{\sss{N_C}})+(\varphi^\dagger\ph^\dagger -b^\ast I_{\sss{N_C}})K^\dagger\Big].
\end{align}
The covariant derivatives are defined by $D_\m \ph = \p_\m\ph-iA_\m\ph$ and $D_\m \varphi = \p_\m \varphi + i\varphi A_\m$.
The Lagrangian has constraints
\begin{align}
&\ph\ph^\dagger-\varphi^\dagger\varphi -c I_{\sss{N_C}} =0,  \label{eq:conts1}\\
&\ph \varphi- b I_{\sss{N_C}}=0,\quad \mbox{(conjugate~transpose)} \label{eq:const2}.
\end{align}
There are two cases for the constraint (\ref{eq:const2}), $b=0$ and $b\neq0$, which are related by the $SU(2)_R$ symmetry.
We consider $b=0$ case in this paper. In this case, field $\ph$ parameterizes the base space of the
Grassmann manifold whereas field $\varphi$ parameterizes the cotangent space. Field $\varphi$ does not contribute to the vacuum
configuration and the BPS solutions of the mass-deformed nonlinear sigma models on the Grassmann manifold \cite{Isozumi:2004jc,Eto:2005cp}.
Then the relevant bosonic part of the Lagrangian (\ref{eq:lag0}) is
\begin{align} \label{eq:lag1}
{\mathcal{L}}=\tr\Big[&D_\m\ph (D^\m\ph)^\dagger-\lt|\ph {\mathcal{M}}-{\mathcal{S}}\ph \rt|^2+\ph\ph^\dagger D-cD\Big].
\end{align}
The Lagrangian has a constraint %
\bea \label{eq:D_term_constraint}
\ph\,\ph^\dagger-cI_{\sss{N_C}}=0.
\eea
By substituting $\mathcal{M}$ and $\mathcal{S}$ in the Lagrangian (\ref{eq:lag1}) with real valued matrices $\wt{M}_\a$ and $\wt{\S}_\a$, $(\a=1,2)$ as
\begin{align}
{\mathcal{M}}=\wt{M}_1+i\wt{M}_2,\quad {\mathcal{S}}=\wt{\S}_1+i\wt{\S}_2,
\end{align}
we get the vacuum condition
\bea \label{eq:vac_cond1}
\ph {\wt{M}_a -\wt{\S}_a} \ph=0, \quad (a=1,2).
\eea
The mass matrices and the real scalar fields can be parameterized as
\begin{align}
&\wt{M}_1=\mathrm{diag}(l_1,l_2,\cdots,l_{\sss{{N_F}}}), \nn\\
&\wt{M}_2=\mathrm{diag}(n_1,n_2,\cdots,n_{\sss{{N_F}}}), \nn\\
&\wt{\S}_1=\mathrm{diag}(\s_1,\s_2,\cdots,\s_{\sss{N_C}}), \nn\\
&\wt{\S}_2=\mathrm{diag}(\ta_1,\ta_2,\cdots,\ta_{\sss{N_C}}).
\end{align}
Then the vacuum solutions are labelled by
\begin{align} \label{eq:vac_sol1}
\lt(\s_1+i\ta_1,\s_2+i\ta_2,\cdots,\s_{\sss{N_C}}+ i\ta_{\sss{N_C}}\rt)=
(l_i+in_i,l_j+in_j,\cdots,l_k+in_k),
\end{align}
where $i,j,k=1,\cdots,N_F$. Therefore there are $_{N_F}C_{N_C}$ vacuum solutions as it is observed in \cite{Arai:2003tc}. The vacuum solutions should be constrained by (\ref{eq:D_term_constraint}).

We are interested in static configurations, which are independent of the $x^3$-coordinate. So we fix $\p_0=\p_3=0$. We also assume that there is the Poincar\'{e} invariance on the worldvolume so we fix $A_0=A_3=0$. Then the energy density is
\begin{align}\label{eq:energy_density}
{\mathcal{E}}&=\tr \lt[\sum_{\a=1,2}\Big| D_\a \ph \mp\lt(\ph \wt{M}_\a - \wt{\S}_\a \ph\rt)\Big|^2 \rt]  \pm {\mathcal{T}} \geq \pm {\mathcal{T}},
\end{align}
where the tension density is
\bea \label{eq:tension_density}
{\mathcal{T}}=\tr\lt[\sum_{\a=1,2}\p_\a\lt(\ph \wt{M}_\a \ph^\dagger\rt)\rt].
\eea
We use the index $\a=1,2$ for codimensions and adjoint scalars as it is done in \cite{Eto:2005cp}. The energy density (\ref{eq:energy_density}) and the tension density (\ref{eq:tension_density}) are constrained by (\ref{eq:D_term_constraint}).

The (anti-)BPS equation is
\bea \label{eq:bps_eq}
D_\a \ph \mp\lt(\ph \wt{M}_\a - \wt{\S}_\a \ph\rt)=0, \quad (\a=1,2).
\eea
We choose the upper sign for the BPS equation and the lower sign for the anti-BPS equation. The BPS solution \cite{Isozumi:2004jc,Eto:2005cp,Eto:2005fm} is
\bea \label{eq:bps_sol}
\ph=S^{-1}H_0e^{\wt{M}_1x^1+\wt{M}_2x^2},
\eea
with a relation
\bea \label{eq:smatrix}
S^{-1}\p_\a S:=\wt{\S}_\a-iA_\a, \quad (\a=1,2).
\eea
The constraint (\ref{eq:D_term_constraint}) becomes
\bea \label{eq:ssd}
SS^\dagger=\frac{1}{c}H_0e^{2\wt{M}_1x^1+2\wt{M}_2x^2}H_0^\dagger.
\eea
The BPS solution (\ref{eq:bps_sol}), $\S_\a$ and $A_\a$ in (\ref{eq:smatrix}) are invariant under the following transformation:
\bea \label{eq:worldvolumesym}
H_0^\prime=VH_0,\quad S^\prime=VS,\quad V\in GL(N_C,\mathbf{C}).
\eea
This equivalent class of $(S,H_0)$ is called worldvolume symmetry in the moduli matrix
formalism \cite{Isozumi:2004jc}. Therefore the moduli space, which is parameterized by $H_0$, is the Grassmann manifold.

$1/1$ BPS supersymmetric vacua and $1/2$ BPS walls can be constructed by a model with real masses and real fields \cite{Isozumi:2004jc}. The $1/4$ BPS junctions are constructed by a model with complex masses and complex fields \cite{Eto:2005cp,Eto:2005fm}.

\section{Vacua and walls}  \label{sec:vac_walls}
\setcounter{equation}{0}
%
We study vacua and walls of the mass-deformed nonlinear sigma models on the Grassmann manifold by using the moduli matrix formalism \cite{Isozumi:2004jc} and the pictorial representation \cite{Lee:2017kaj,Shin:2018chr}. We can simplify the Lagrangian (\ref{eq:lag1}) by introducing a real mass matrix and a real scalar matrix field. The vacuum condition is
\bea
\ph M -\S\ph=0.
\eea
$M$ and $\S$ are related to $\mathcal{M}$ and $\mathcal{S}$ in (\ref{eq:lag1}) \cite{Eto:2005cp} by
\begin{align}
&{\mathcal{M}}=e^{i\th}(M-\D M I_{N_F}), \nn\\
&{\mathcal{S}}=e^{i\th}(\S-\D M I_{N_C}).
\end{align}
The matrices can be parameterized as
\begin{align}
&M=\mathrm{diag}(m_1,m_2,\cdots,m_{N_F}), \nn\\
&\S=\mathrm{diag}(\s_1,\s_2,\cdots,\s_{\sss{N_C}}).
\end{align}
We can set $m_1>m_2>\cdots>m_N$ without loss of generality since we are interested in generic mass parameters. Then the vacuum solutions are labelled by
\bea \label{eq:vac_sol12}
(\s_1,\s_2,\cdots,\s_{N_C})=(m_i,m_j,\cdots,m_k),
\eea
where $i,j,k=1,\cdots,N_F$. The vacuum solutions are the same as the ones labelled by (\ref{eq:vac_sol1}).

We study walls. We can assume that fields are static and all the fields depend on $x_1\equiv x$ coordinate. We can also assume that there is the Poincar\'{e} invariance on the worldvolume of walls so we set $A_0=A_2=A_3=0$. We can learn from (\ref{eq:bps_sol}) that the BPS solution \cite{Isozumi:2004jc} is
\bea
\ph=S^{-1}H_{0}e^{Mx}.
\eea

Walls are constructed from elementary walls in the moduli matrix formalism. Let $\la A \ra$ denote a vacuum and $\la A \leftarrow B\ra$ denote a wall that connects vacuum $\la A \ra$ and vacuum $\la B \ra$. The moduli matrix of elementary wall $\la A \leftarrow B\ra$ is
\begin{align} \label{eq:elemwall}
&H_{0\la A \leftarrow B\ra}=H_{0\la A \ra} e^{E_i(r)}, \nn\\
&E_i(r)\equiv e^r E_i,\quad (i=1,\cdots,N),
\end{align}
where $E_i$ is an elementary wall operator and $r$ is a complex parameter with $-\infty$ $<$ $\mathrm{Re}(r)$$<$$+\infty$. The elementary wall operator $E_i$ is a simple root generator of $SU(N_F)$,
which satisfies
\bea \label{eq:stepop}
c[M,E_i]=c(\vec{m}\cdot \vec{a}_i)E_i=T_{\la A \leftarrow B\ra} E_i,
\eea
where $c$ is the electric FI parameter of the Lagrangian (\ref{eq:n=2_lag}), $\vec{m}=(m_1,\cdots,m_{\sss{N_F}})$, and $T_{\la A \leftarrow B\ra}$ is the tension of the elementary wall. Elementary walls can be identified with the simple roots of $SU(N_F)$ \cite{Sakai:2005sp}.

The simple root generators and the simple roots of $SU(N)$ \cite{Isaev:2018xcg} are
\bea \label{eq:simple_roots}
&&E_i=e_{i,i+1}, \nn\\
&&\vec{\a}_i=\hat{e}_i-\hat{e}_{i+1},\quad (i=1,\cdots,N-1).
\eea
The set of vectors $\{\hat{e}_i\}$ is the orthogonal unit vectors $\hat{e}_i\cdot\hat{e}_j=\d_{ij}$.

Elementary walls can be compressed to single walls. A compressed wall of level $n$ which connects $\la A \ra$ and $\la A^\prime \ra$ is
\begin{align} \label{eq:compwall}
&H_{0\la A \leftarrow A^\prime \ra}=H_{0\la A \ra}e^{[E_{i_1},[E_{i_2},[E_{i_3},[\cdots,[E_{i_n},E_{i_{n+1}}]\cdots]]](r)}, \nn\\
&(i_m=1,\cdots,N;~m=1,\cdots,n+1).
\end{align}
A multiwall is constructed by multiplying a single wall operator to another wall moduli matrix. A multiwall interpolating $\la
A \ra$, $\la A^\prime \ra$,$\cdots$, and $\la B \ra$ is
\begin{align}
&H_{0\lra{A \larw A^\prime \larw \cdots \larw B}}=H_{0\lra{A}}e^{E_{i_1}(r_1)}e^{E_{i_2}(r_2)}\cdots
e^{E_{i_n}(r_n)}, \nn\\
&(i_m=1,\cdots,N;~m=1,\cdots,n),
\end{align}
where parameters $r_i$, $(i=1,2,\cdots)$ are complex parameters with $-\infty$$<$$\mathrm{Re}(r_i)$$<$$+\infty$. Penetrable walls pass through each other since the wall operators commute:
\bea \label{eq:penetwall}
[E_{i_m},E_{i_n}]=0.
\eea

Let vector $\vec{g}_{\lra{A\larw A^\prime}}$ denote the wall that interpolates vacuum $\lra{A}$ and vacuum $\lra{A^\prime}$. Then the elementary wall (\ref{eq:elemwall}) with the relation (\ref{eq:stepop}) can be identified with
\bea
\vec{g}_{\lra{A\larw B}}\equiv c\vec{\a}_i.
\eea
The tension of the wall can be read from (\ref{eq:stepop}) as
\bea
T_{\lra{A\larw A^\prime}}=\vec{m}\cdot \vec{g}_{\lra{A\larw A^\prime}}.
\eea
The compressed wall in (\ref{eq:compwall}) is identified with
\bea
\vec{g}_{\lra{A\larw A^\prime}}\equiv c\vec{\a}_{i_1}+c\vec{\a}_{i_2}+ c\vec{\a}_{i_3}+\cdots + c\vec{\a}_{i_n}+ c\vec{\a}_{i_{n+1}}.
\eea
The root vectors of the two penetrable walls of (\ref{eq:penetwall}) are orthogonal
\bea
\vec{g}_{i_m}\cdot \vec{g}_{i_n}=0.
\eea

\begin{figure}[ht!]
\begin{center}
\bea
\begin{array}{ccc}
\includegraphics[width=5cm,clip]{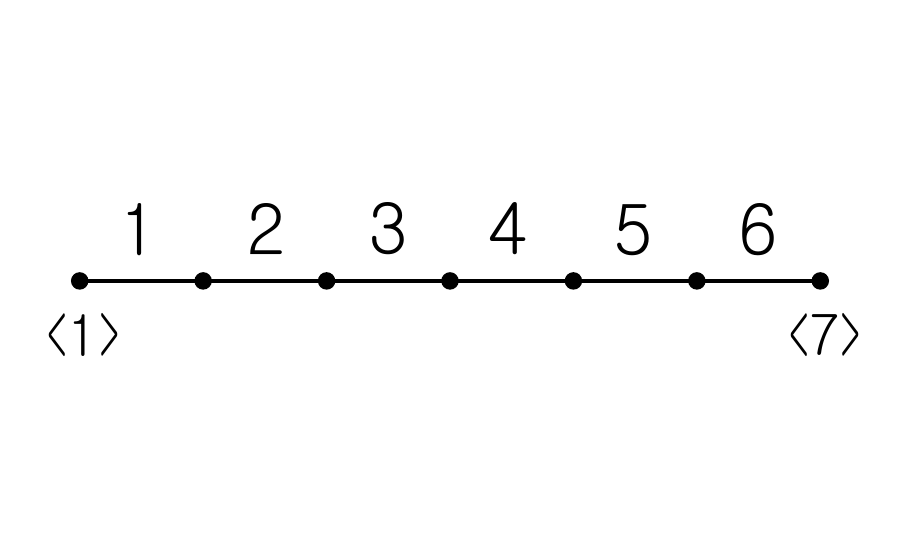}
&~~&
\includegraphics[width=6cm,clip]{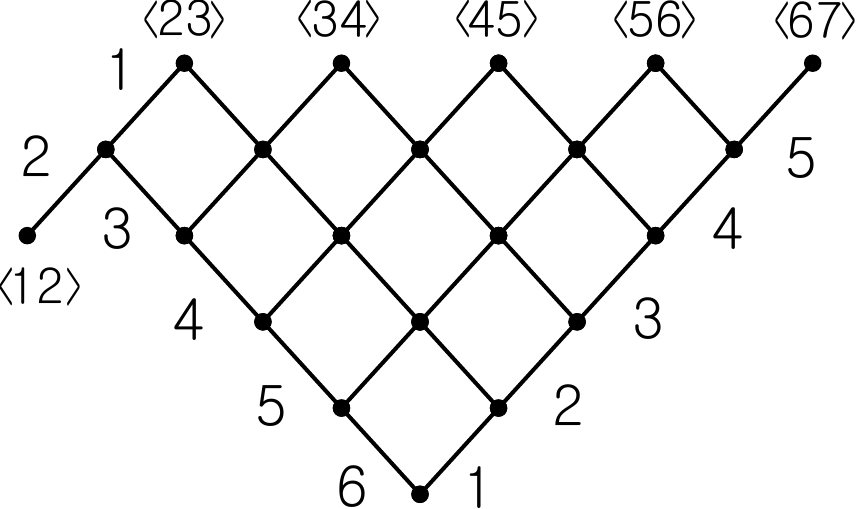}\\
\mathrm{(a)} &~~~~~~& \mathrm{(b)}
\end{array}
\eea
\includegraphics[width=9cm,clip]{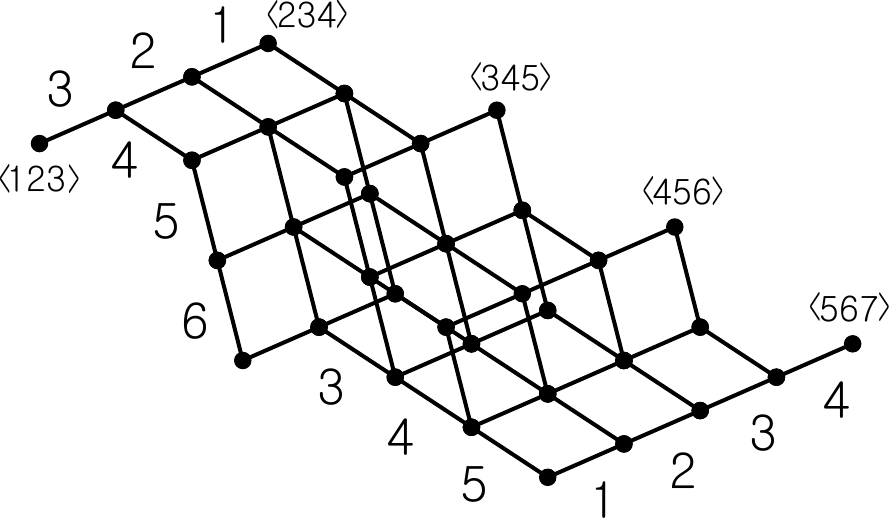}\\
(c)
\end{center}
 \caption{ Vacua and elementary walls in $G_{7,N}$. The numbers indicate the subscript $i$'s of simple roots $\vec{\a}_i$. (a)$G_{7,1}$ (b)$G_{7,2}$ (c) $G_{7,3}$. }
 \label{fig:g7n}
\end{figure}

The pictorial representation of \cite{Lee:2017kaj} is applied to vacua and walls of mass-deformed nonlinear sigma
models on the Grassmann manifold in \cite{Shin:2018chr}. In the representation, vacua and elementary walls are
described by vertices and segments. It is observed that the duality $N_C \leftrightarrow N_F-N_C$ corresponds to a
$\pi$-rotation in the pictorial representation and the diagram of $G_{N_F+1,N_C}$ repeats the diagram of $G_{N_F,N_C}$. The
diagrams of the vacua and the elementary walls of the mass-deformed nonlinear sigma models on $G_{N_F,N_C}$, with
$(N_F,N_C)$$=$$(4,2)$,$(5,2)$,$(5,3)$,$(6,2)$,$(6,3)$,$(6,4)$ are shown in \cite{Shin:2018chr}. The diagrams of the vacua and
the elementary walls of the mass-deformed nonlinear sigma models on $G_{7,N_C}$ with $N_C$$=$$1$,$2$,$3$ are presented in Figure
\ref{fig:g7n}.

The mass-deformed nonlinear sigma models on $G_{N_F,N_C}$ have $_{N_F}C_{N_C}$ vacua. All the walls of the model are constructed from elementary walls, which can be described by the simple roots of $SU(N_F)$, $\vec{\a}_i$, ($i=1,\cdots,N_F-1$) of (\ref{eq:simple_roots}). As we can see in \cite{Shin:2018chr} and Figure \ref{fig:g7n}, the diagrams for $G_{N_F,N_C}$ repeats the diagrams for $G_{n,N_C}$, $(N_C<n\leq N_F-1)$. It shows that vacua and elementary walls of the mass-deformed nonlinear sigma models on $G_{N_F+1,N_C}$ can be derived from the configuration of vacua and elementary walls of $G_{N_F,N_C}$, by adding $_{N_F}C_{N_C-1}$ number of vacua and elementary walls. We can focus on the case where $N_C\leq \lt[\frac{N_F}{2}\rt]$ since the case with $N_C>\lt[\frac{N_F}{2}\rt]$ can be determined by the duality condition $G_{N_F,N_C}\simeq G_{N_F,N_F-N_C}$, which is related by a $\pi$-rotation in the pictorial representation.

\begin{figure}[ht!]
\begin{center}
\bea
\begin{array}{ccc}
\includegraphics[width=6cm,clip]{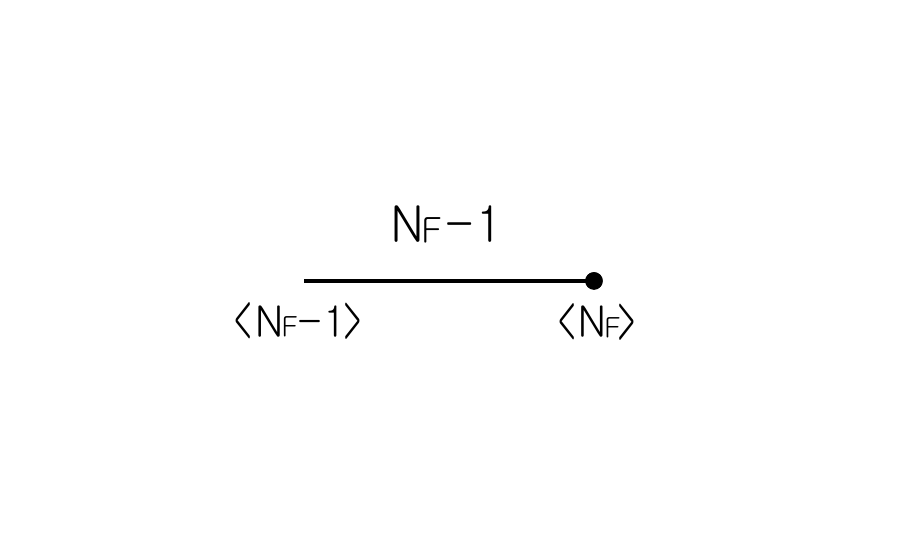}
&~~&
\includegraphics[width=6cm,clip]{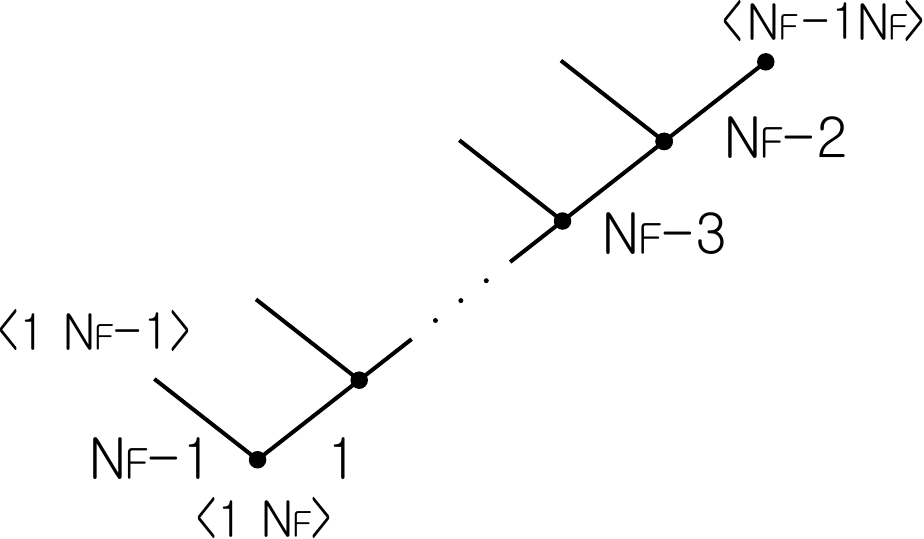}\\
\mathrm{(a)} &~~~~& \mathrm{(b)}\\
~ & ~ & ~ \\
~ & ~ & ~
\end{array}
\eea
\includegraphics[width=12cm,clip]{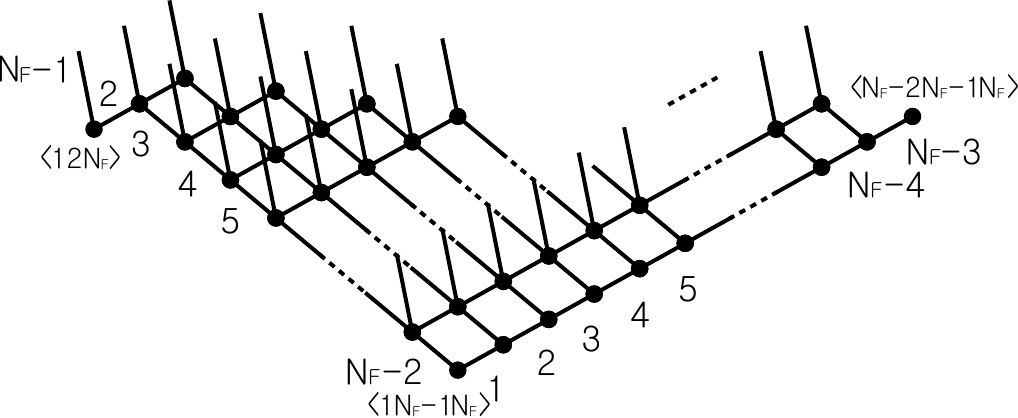}\\
(c)
\end{center}
 \caption{ (a)$N_C=1$ ~(b)$N_C=2$ ~(c)$N_C=3$ }
 \label{fig:nc1nc2nc3}
\end{figure}

We present diagrams for $N_C$$=$$1$,$2$,$3$ in Figure \ref{fig:nc1nc2nc3}. The structure of vacua and elementary walls of the mass-deformed nonlinear sigma models on $G_{N_F,1}$ is constructed by adding one vacuum $\lra{N_F}$ and one elementary wall $\vec{\a}_{N_F-1}$ to the structure of $G_{N_F-1,1}$. The diagram of the additional structure is depicted in Figure \ref{fig:nc1nc2nc3}(a). The structure of vacua and elementary walls of the mass-deformed nonlinear sigma models on $G_{N_F,2}$ is constructed by adding the diagram in Figure \ref{fig:nc1nc2nc3}(b), which consists of $(N_F-1)$ vacua, $(N_F-2)$ number of $\vec{\a}_{N_F-1}$ and one for each $\vec{\a}_i$, $i=1,\cdots,N_F-2$, to the diagram for $G_{N_F-1,2}$. In the similar manner, the diagram of vacua and elementary walls of mass-deformed nonlinear sigma models on $G_{N_F,3}$ is obtained by adding the diagram in Figure \ref{fig:nc1nc2nc3}(c) to the diagram for $G_{N_F-1,3}$. The diagram adds $(N_F-3)$ number of each $\vec{\a}_i$ $(i=1,\cdots,N_F-2)$ and $\frac{(N_F-3)(N_F-2)}{2}$ number of $\vec{\a}_{N_F-1}$. The configuration of vacua and walls of the mass-deformed nonlinear sigma models on $G_{N_F,N_C}$ can be systematically derived from the diagrams of the pictorial representation.

We have discussed generic properties of vacua and elementary walls of the nonlinear sigma models on the Grassmann manifold with flavour number $N_F$ for colour number $N_C=1,2,3$ as examples. Non-orthogonal elementary walls are aligned next to each other. Orthogonal elementary walls form parallelograms. Since the global symmetry is $SU(N_F)$, diagrams for $N_C\geq 4$ cases are trivial extensions. For a finite number of discrete vacua, we can always identify the simple roots of $SU(N_F)$ that interpolate the vacua. We present the diagram for vacua and elementary walls of the mass-deformed nonlinear sigma model on $G_{8,4}$ as an example in \ref{app:g84}.

\section{Three-pronged junctions } \label{sec:junctions}
\setcounter{equation}{0}

A three-pronged junction is formed by three vacua interpolated by three non-penetrable walls. A wall that interpolates vacuum $\lra{A}$ and vacuum $\lra{B}$ has a tension $\vec{T}^{AB}$ pulling the junction along the wall direction outwards \cite{Eto:2005cp}
\begin{align}
&\vec{T}^{\lra{AB}}=(-Z_2^{AB},Z_1^{AB}),\nn\\
&(Z_1^{AB},Z_2^{AB})\equiv c(l_A-l_B,n_A-n_B),
\end{align}
which is computed by the line integral of the tension density defined by (\ref{eq:tension_density}).

\begin{figure}[ht!]
\begin{center}
\includegraphics[width=10cm,clip]{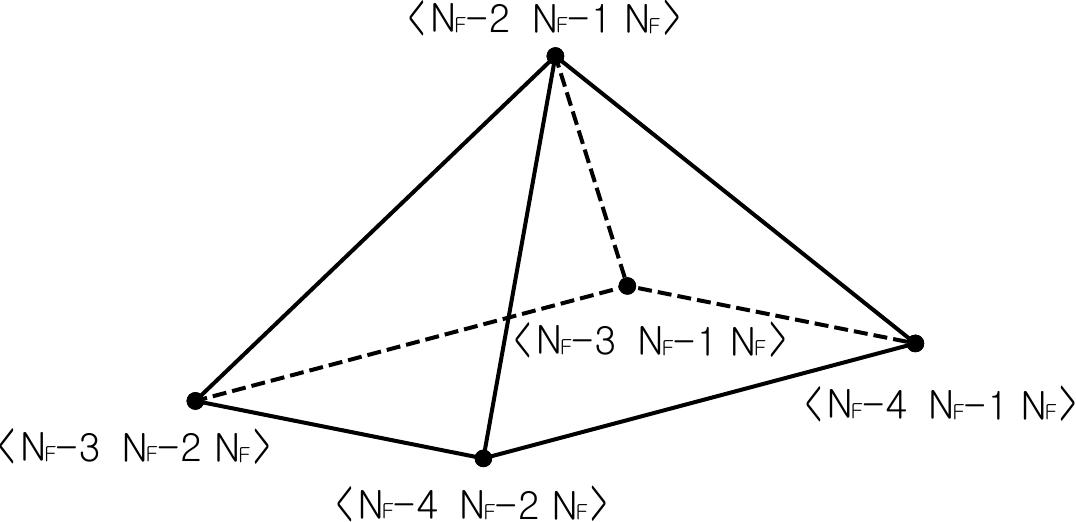}\\
\end{center}
 \caption{ Vertices, edges and triangular faces correspond to vacua, walls and three-pronged junctions. The parallelogram shaped base corresponds to two pairs of penetrable walls. }
 \label{fig:pyramid}
\end{figure}

We study three-pronged junctions of the mass-deformed nonlinear sigma models on the Grassmann manifold $G_{N_F,3}$. We reformulate the
diagram in Figure \ref{fig:nc1nc2nc3}(c) by connecting adjacent vacua to produce a pyramid of which the vertices, the edges, the triangular faces and the parallelogram shaped base correspond to vacua, walls, three-pronged junctions and two pairs of penetrable walls \cite{Shin:2018chr}\footnote{This type of three-dimensional diagram is applied to $\mathbf{C}P^3$ in \cite{Eto:2005cp}.}. The pyramid
is depicted in Figure \ref{fig:pyramid}. Let us use semicolons in vacuum labels $\lra{i;j;\cdots;k}$ to separate the flavour numbers.

There are two types of junctions \cite{Eto:2005fm}. Abelian junctions divide a set of three vacua with labels that differ by one components: $\lra{\cdots,A}$, $\lra{\cdots,B}$, $\lra{\cdots,C}$. Non-Abelian junctions divide a set of three vacua with labels that differ by two components: $\lra{\cdots,AB}$, $\lra{\cdots,BC}$, $\lra{\cdots,AC}$. Abelian junctions exist in Abelian gauge theories and non-Abelian gauge theories whereas non-Abelian junctions exist only in non-Abelian gauge theories.

There are two Abelian junctions and two non-Abelian junctions in Figure \ref{fig:pyramid}:
\begin{align}
\bullet&\mathrm{abelian~junctions} \nn\\
&\Big\{ \lra{N_F-4 ; N_F-2; N_F}, \lra{N_F-3;N_F-2;N_F}, \lra{N_F-2;N_F-1; N_F}\Big\}, \nn\\
&\Big\{ \lra{N_F-4;N_F-1;N_F}, \lra{N_F-3;N_F-1;N_F}, \lra{N_F-2;N_F-1;N_F}\Big\}. \label{ea:abel_junction} \\
&~~\nn\\
\bullet&\mbox{non-abelian junctions} \nn\\
&\Big\{ \lra{N_F-4;N_F-2;N_F}, \lra{N_F-4;N_F-1;N_F}, \lra{N_F-2;N_F-1;N_F}\Big\}, \nn\\
&\Big\{ \lra{N_F-3;N_F-1;N_F}, \lra{N_F-3;N_F-2;N_F}, \lra{N_F-2;N_F-1;N_F}\Big\}. \label{ea:nonabel_junction}
\end{align}

We study Abelian junction $\Big\{\la N_F-4 ; N_F-2 ; N_F\ra$,$\la N_F-3; N_F-2;
N_F\ra$,$\la N_F-2;N_F-1;N_F\ra \Big\}$. The moduli matrix for the junction is
\begin{align} \label{eq:moduli_abel}
&H_0=\lt(
\begin{array}{ccccccc}
\cdots  &  h_{1\,{\sss{N_F}}-4}  &  h_{1\,{\sss{N_F}}-3}  &  0  &  h_{1\,{\sss{N_F}}-1}  &  0 \\
\cdots  &  0 &  0  &  h_{2\,{\sss{N_F}}-2} &  0  &  0   \\
\cdots  &  0 &  0  &  0  &  0  &  h_{3\,{\sss{N_F}}}
\end{array}
\rt), \nn\\
&h_{ai}:=\exp(a_{ai}+ib_{ai}).
\end{align}
The limit of (\ref{eq:moduli_abel}) as $a_{1\,{\sss{N_F-1}}}\rightarrow -\infty$ is the elementary wall that interpolates $\lra{N_F-4 ; N_F-2 ; N_F}$ and $\lra{N_F-3; N_F-2; N_F}$. The limit of (\ref{eq:moduli_abel}) as $a_{1\,{\sss{N_F-1}}}\rightarrow -\infty$ and $a_{1\,{\sss{N_F-3}}}\rightarrow -\infty$, keeping $a_{1\,{\sss{N_F-4}}}$ finite is $\lra{N_F-4 ; N_F-2 ; N_F}$. The limit of (\ref{eq:moduli_abel}) as $a_{1\,{\sss{N_F-1}}}\rightarrow -\infty$ and $a_{1\,{\sss{N_F-4}}}\rightarrow -\infty$, keeping $a_{1\,{\sss{N_F-3}}}$ finite is $\lra{N_F-3 ; N_F-2 ; N_F}$. 
Equation (\ref{eq:ssd}) with (\ref{eq:moduli_abel}) is computed as
\begin{align}\label{eq:ssd1}
&SS^\dagger=\frac{1}{c}H_0e^{(2\wt{M}_1 x^1+2\wt{M}_2 x^2)}H_0^\dagger=\frac{1}{c}\mathrm{diag}(\D_1,\D_2,\D_3), \nn\\
&\D_1=\exp(2l_{{\sss{N_F}}-4}x^1+2n_{{\sss{N_F}}-4}x^2+2a_{1\,{{\sss{N_F}}-4}}) + \exp(2l_{{\sss{N_F}}-3}x^1+2n_{{\sss{N_F}}-3}x^2+2a_{1\,{{\sss{N_F}}-3}})  \nn\\
&~\quad+\exp(2l_{{\sss{N_F}}-1}x^1+2n_{{\sss{N_F}}-1}x^2+2a_{1\,{\sss{N_F}}-1}),   \nn\\
&\D_2=\exp(2l_{{\sss{N_F}}-2}x^1+2n_{{\sss{N_F}}-2}x^2+2a_{2\,{\sss{N_F}}-2}), \nn\\
&\D_3=\exp(2l_{\sss{N_F}}x^1+2n_{\sss{N_F}}x^2+2a_{3\,{\sss{N_F}}}).
\end{align}
The BPS solution (\ref{eq:bps_sol}) with (\ref{eq:moduli_abel}) and (\ref{eq:ssd1}) is
\begin{align}\label{eq:abel_junc_sol}
\ph&=S^{-1}H_0e^{\wt{M}_1 x^1+\wt{M}_2 x^2} \nn\\
&=\sqrt{c}\lt(
\begin{array}{ccccccc}
\cdots   &  \frac{f_{1\,{\nf-4}}}{\sqrt{\D_1}}  & \frac{f_{1\,{\nf-3}}}{\sqrt{\D_1}}  &  0  &  \frac{f_{1\,{\nf-1}}}{\sqrt{\D_1}}  &  0 \\
\cdots   &  0 &  0  &  \frac{f_{2\,{\nf-2}}}{\sqrt{\D_2}}   &  0  &  0   \\
\cdots   &  0 &  0  &  0 &  0  &  \frac{f_{3\,{\sss{N_F}}}}{\sqrt{\D_3}}
\end{array}
\rt), \nn\\
&f_{ai}:=\exp(l_i x^1+n_i x^2 +a_{ai}+ib_{ai}).
\end{align}
Component $(2,N_F-2)$ and component $(3,N_F)$ of (\ref{eq:abel_junc_sol}) are $\sqrt{c}$ with $U(1)$ phases. Since the moduli matrix (\ref{eq:moduli_abel}) has the worldvolume symmetry (\ref{eq:worldvolumesym}), only two of the parameters, $h_{1i}$, $(i=N_F-4,N_F-3,N_F-1)$ are independent. It is consistent with the fact that moduli matrix (\ref{eq:moduli_abel}) consists of three walls and the position of the junction is determined by two of them.

The position of the junction is the solution of the equations:
\begin{align}
{\mathrm{Re}}(f_{1\,\nf-4})={\mathrm{Re}}(f_{1\,\nf-3})={\mathrm{Re}}(f_{1\,\nf-1}),
\end{align}
which produce
\begin{align}
&(l_{{\sss{N_F}}-4}-l_{{\sss{N_F}}-3})x^1+(n_{{\sss{N_F}}-4}-n_{{\sss{N_F}}-3})x^2+a_{1\,{\sss{N_F}}-4}-a_{1\,{\sss{N_F}}-3}=0,  \nn\\
&(l_{{\sss{N_F}}-3}-l_{{\sss{N_F}}-1})x^1+(n_{{\sss{N_F}}-3}-n_{{\sss{N_F}}-1})x^2+a_{1\,{\sss{N_F}}-3}-a_{1\,{\sss{N_F}}-1}=0.
\end{align}
The position of the junction is
\bea
(x^1,x^2)=\lt(\frac{S_1}{S_3},\frac{S_2}{S_3}\rt),
\eea
\begin{align}
S_1=&(-n_{\nf-3}+n_{\nf-1})a_{1 \,\nf-4} + (-n_{\nf-1}+n_{\nf-4})a_{1\, \nf-3} \nn\\
&+(-n_{\nf-4}+n_{\nf-3})a_{1\, \nf-1}, \nn\\
S_2=&(l_{\nf-3}-l_{\nf-1})a_{1\,\nf-4} + (-l_{\nf-4}+l_{\nf-1})a_{1\,\nf-3} \nn\\
&+(l_{\nf-4}-l_{\nf-3})a_{1\,\nf-1}, \nn\\
S_3=&(-l_{\nf-3}+l_{\nf-1})n_{\nf-4}+(-l_{\nf-1}+l_{\nf-4})n_{\nf-3} \nn\\
&+(-l_{\nf-4}+l_{\nf-3})n_{\nf-1}.
\end{align}
As mentioned previously, only two of the parameters, $a_{1i}$, $(i=N_F-4,N_F-3,N_F-1)$ are independent.

The matrix $SS^\dagger$ defined by (\ref{eq:ssd}) is not diagonal for non-Abelian three-pronged junctions as the vacua differ by two label components. A detour is needed to apply the moduli matrix formalism to junctions. In \cite{Eto:2005fm}, the Grassmann manifold is embedded into the complex projective space by the Pl\"{u}cker embedding. In \cite{Shin:2018chr}, the moduli matrix formalism is applied to each wall that form a three-pronged junction as the set of walls can be identified in the diagrams of the pictorial representation.

We study non-Abelian three-pronged junction $\Big\{\la N_F-4 ; N_F-2 ; N_F\ra,~\la N_F-4 ; N_F-1 ; N_F\ra ,~\la N_F-2 ; N_F-1
; N_F \ra \Big\}$ following the method of \cite{Shin:2018chr}. The moduli matrix of the wall, which interpolates $\la N_F-4 ; N_F-2 ; N_F \ra$ and $\la N_F-4 ; N_F-1 ; N_F \ra$ is
\begin{align} \label{eq:nonabeljun_modul_1}
&H_0=\lt(
\begin{array}{ccccccc}
\cdots  &    h_{1\,\nf-4}  &  0  &  0  &  0  &  0 \\
\cdots  &    0 &  0  & h_{2\,\nf-2}   &  h_{2\,\nf-1} &  0   \\
\cdots  &    0 &  0  &  0  &  0  &  h_{3\,\nf}
\end{array}
\rt), \nn\\
&h_{ai}:=\exp(a_{ai}+ib_{ai}).
\end{align}
The wall solution is
\begin{align} \label{eq:nonabeljun1}
\ph&=S^{-1}H_0e^{\wt{M}_1 x^1+\wt{M}_2 x^2} \nn\\
&=\sqrt{c}\lt(
\begin{array}{ccccccc}
\cdots  &    \frac{f_{1\,\nf-4}}{\sqrt{\D_1}}  &  0  &  0 &  0  &  0 \\
\cdots  &    0 &  0  &  \frac{f_{2\,\nf-2}}{\sqrt{\D_2}}  &  \frac{f_{2\,\nf-1}}{\sqrt{\D_2}}  &  0   \\
\cdots  &    0 &  0  &  0  &  0  &  \frac{f_{3\,\nf}}{\sqrt{\D_3}}
\end{array}
\rt), \nn\\ ~\nn\\
&f_{ai}:=\exp(l_i x^1+n_i x^2+a_{ai}+ib_{ai}), \nn\\  ~ \nn\\
&\D_1=\exp(2l_{{\sss{N_F}}-4}x^1+2n_{{\sss{N_F}}-4}x^2 + 2a_{1{\sss{N_F}}-4}), \nn\\
&\D_2=\exp(2l_{{\sss{N_F}}-2}x^1+2n_{{\sss{N_F}}-2}x^2 + 2a_{2{\sss{N_F}}-2}) \nn\\ 
&\quad \quad + \exp(2l_{{\sss{N_F}}-1}x^1+2n_{{\sss{N_F}}-1}x^2 + 2a_{2{\sss{N_F}}-1}), \nn\\
&\D_3=\exp(2l_{{\sss{N_F}}}x^1+2n_{{\sss{N_F}}}x^2 + 2a_{3{\sss{N_F}}}).
\end{align}
Component $(1,N_F-4)$ and component $(3,N_F)$ of (\ref{eq:nonabeljun1}) are $\sqrt{c}$ with $U(1)$ phases. Since moduli matrix (\ref{eq:nonabeljun_modul_1}) has the worldvolume symmetry (\ref{eq:worldvolumesym}), only one of the parameters, $h_{2i}$, $(i=N_F-2,N_F-1)$ is independent. The wall is located in the region where ${\mathrm{Re}}(f_{2\,\nf-2})={\mathrm{Re}}(f_{2\,\nf-1})$:
\bea \label{eq:nawall_1}
(l_{\nf-2}-l_{\nf-1})x^1+(n_{\nf-2}-n_{\nf-1})x^2+(a_{2\,\nf-2}-a_{2\,\nf-1})=0.
\eea
Only one of the parameters, $a_{2i}$, $(i=N_F-2,N_F-1)$ is independent.

The moduli matrix of the wall that interpolates $\la N_F-4 ; N_F-1 ; N_F \ra$ and $\la N_F-2 ; N_F-1 ; N_F \ra$ is
\begin{align}
H_0=\lt(
\begin{array}{ccccccc}
\cdots  &    h_{1\,\nf-4}  &  0  &  h_{1\,\nf-2}  &  0  &  0 \\
\cdots  &    0 &  0  &  0 &  h_{2\,\nf-1}  &  0   \\
\cdots  &    0 &  0  &  0  &  0  &  h_{3\,\nf}
\end{array}
\rt), \quad
h_{ai}:=e^{(a_{ai}+ib_{ai})},
\end{align}
and the position of the wall is
\bea \label{eq:nawall_2}
(l_{\nf-4}-l_{\nf-2})x^1+(n_{\nf-4}-n_{\nf-2})x^2+a_{1\,\nf-4}-a_{1\,\nf-2}=0.
\eea
The moduli matrix of the wall that interpolates $\la N_F-2 ; N_F-1 ; N_F \ra$ and $\la N_F-4 ; N_F-2 ; N_F \ra$ is
\begin{align}
H_0=\lt(
\begin{array}{ccccccc}
\cdots  &    0  &  0  &  h_{1\,\nf-2} &  0  &  0 \\
\cdots  &    h_{2\,\nf-4} &  0  &  0 &  h_{2\,\nf-1}  &  0   \\
\cdots  &    0 &  0  &  0  &  0  &  h_{3\,\nf}
\end{array}
\rt),\quad
h_{ai}:=e^{(a_{ai}+ib_{ai})},
\end{align}
and the position of the wall is
\bea \label{eq:nawall_3}
(l_{\nf-4}-l_{\nf-1})x^1+(n_{\nf-4}-n_{\nf-1})x^2+(a_{2\,\nf-4}-a_{2\,\nf-1})=0.
\eea

The position of the junction can be calculated by solving the equations (\ref{eq:nawall_1}), (\ref{eq:nawall_2}) and (\ref{eq:nawall_3}). The consistency condition is
\bea
a_{1\,\nf-4}-a_{1\,\nf-2}=a_{2\,\nf-4}-a_{2\, \nf-2}.
\eea
Therefore there are two independent parameters for the three-pronged junction as expected. 
The position of the junction is
\begin{align}
(x^1,x^2)=\lt(\frac{T_1}{T_3},\frac{T_2}{T_3}\rt),
\end{align}
\begin{align}
T_1=&(n_{\nf-2}-n_{\nf-1})a_{2\,\nf-4}+(n_{\nf-1}-n_{\nf-4})a_{2\,\nf-2} \nn\\
    &+(n_{\nf-4}-n_{\nf-2})a_{2\,\nf-1},  \nn\\
T_2=&(-l_{\nf-2}+l_{\nf-1})a_{2\,\nf-4}+(-l_{\nf-1}+l_{\nf-4})a_{2\,\nf-2} \nn\\
    &+(-l_{\nf-4}+l_{\nf-2})a_{2\,\nf-1},  \nn\\
T_3=&(l_{\nf-2}-l_{\nf-1})n_{\nf-4}+(l_{\nf-1}-l_{\nf-4})n_{\nf-2} \nn\\
    &+(l_{\nf-4}-l_{\nf-2})n_{\nf-1}.
\end{align}

We have computed junction positions of an Abelian three-pronged junction and a non-Abelian three-pronged junction in the pictorial representation by using $N_C\times N_F$ moduli matrices instead of using the Pl\"{u}cker embedding.

\section{Summary and discussion}  \label{sec:summary}
\setcounter{equation}{0}
We have discussed vacua and BPS objects of the $\mathcal{N}=2$ mass-deformed nonlinear sigma models on the Grassmann manifold $G_{N_F,N_C}$ with $N_C=1,2,3$ as an extension of \cite{Shin:2018chr}. We have applied the moduli matrix formalism \cite{Isozumi:2004jc} and the pictorial representation \cite{Lee:2017kaj,Shin:2018chr} to vacua, walls and three-pronged junctions. Since we can analyse three-pronged junctions with $N_C\times N_F$ moduli matrices in the pictorial representation, we can apply the moduli matrix formalism to three-pronged junctions of the nonlinear sigma models on $SO(2N)/U(N)$ and $Sp(N)/U(N)$, which are quadrics of the Grassmann manifold. We hope to report on the results elsewhere.

We have considered the mass-deformed hyper-K\"{a}hler nonlinear sigma models with eight supercharges on the Grassmann manifold in four dimensions. The target space of hypermultiplets for locally supersymmetric nonlinear sigma models with eight supercharges in the dimension d$\leq$6 are quaternionic K\"{a}hler manifolds \cite{Bagger:1983tt}. Exact BPS domain wall solutions are studied in five dimensional supergravity coupled with the hypermultiplets and vector multiplet \cite{Arai:2002ph}. It is worth considering whether we can improve the method proposed in this work to apply it to the supergravity models. \\~\\

\noindent {\bf Acknowledgement}

The author would like to thank A. Golubtsova for early participation and D. Bykov for helpful comments. The author also would like to thank  Bogoliubov Laboratory of Theoretical Physics, Joint Institute for Nuclear Research for warm hospitality. This research was supported by Basic Science Research Program through the National Research Foundation of Korea (NRF-2017R1D1A1B03034222).



\appendix
\def\thesection{Appendix \Alph{section}}
\setcounter{equation}{0}
\renewcommand{\theequation}{\Alph{section}.\arabic{equation}}


\section{Diagram for vacua and elementary walls of the mass-deformed nonlinear sigma models on $G_{8,4}$}  \label{app:g84}
\begin{figure}[ht!]
\begin{center}
\includegraphics[width=13cm,clip]{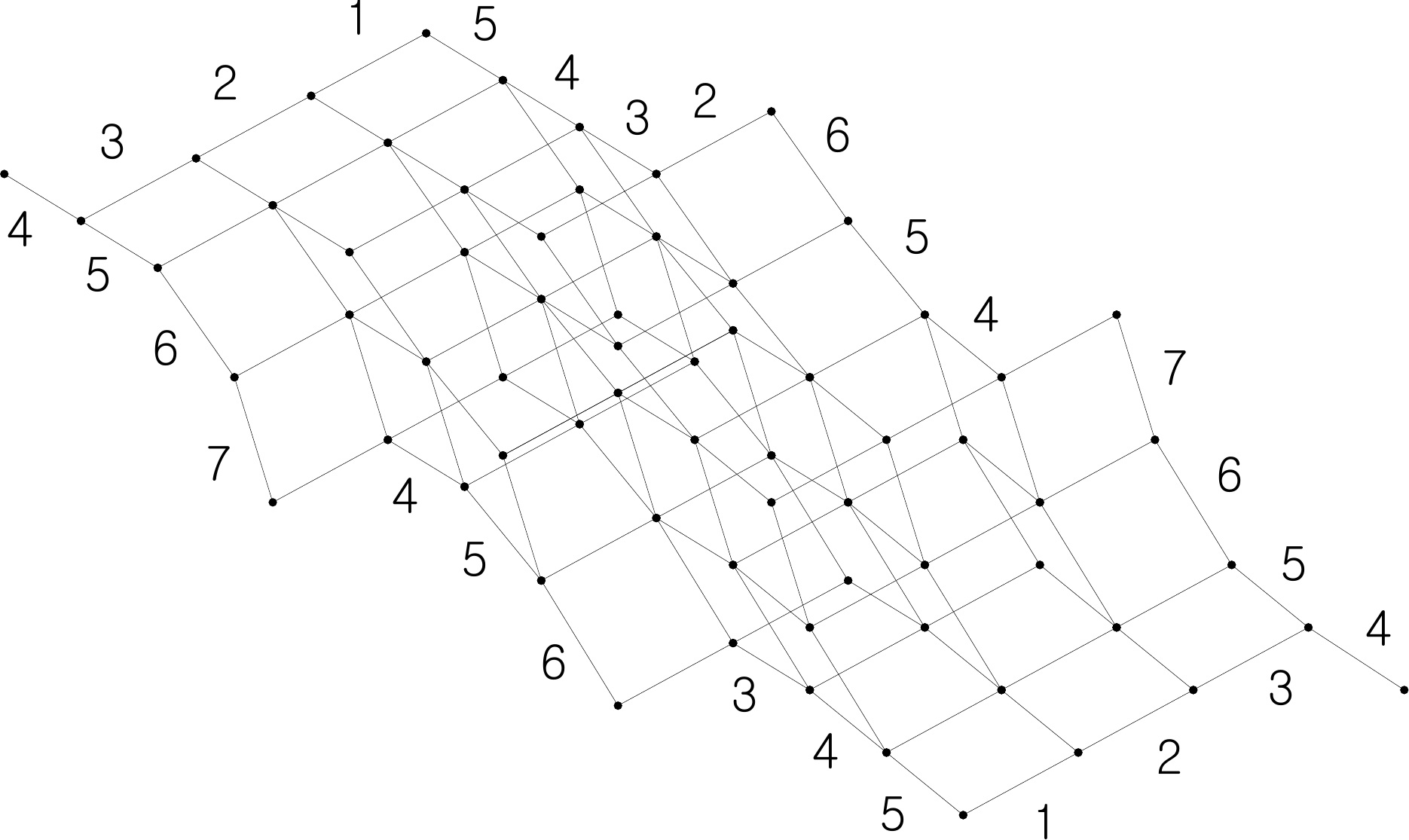}\\
\end{center}
 \caption{Vacua and elementary walls in $G_{8,4}$. The numbers indicate the subscript $i$'s of simple roots $\vec{\a}_i$. }
 \label{fig:g84}
\end{figure}

\setcounter{equation}{0}
We present the diagram for vacua and elementary walls of the mass-deformed nonlinear sigma models on the Grassmann manifold $G_{8,4}$ in Figure \ref{fig:g84}. The manifold $G_{8,4}$ is self-dual under $N_F\leftrightarrow N_F-N_C$. Therefore the diagram has a $\pi$-rotation symmetry.

In Figure \ref{fig:g7n}(c), we have presented the diagram for $G_{7,3}$. The diagram for $G_{7,4}$ is obtained by a $\pi$-rotation of the diagram in Figure \ref{fig:g7n}(c). Non-orthogonal elementary walls are aligned next to each other while orthogonal elementary walls form parallelograms. By adding $\a_7$ vectors to proper positions of the diagram for $G_{7,4}$, we can build the diagram for $G_{8,4}$, which is presented in Figure \ref{fig:g84}. \\~\\

\end{document}